\documentstyle[emulateapj,apjfonts,psfig]{article}

\lefthead{J.~M.~Miller et al.}  
\righthead{{\it XMM-Newton} Observations of GX 339$-$4}

\received{Date}
\revised{Date}
\accepted{Date}

\journalid{vol}{date}
\articleid{1}{4}
\paperid{id}

\cpright{AAS}{1999}
\ccc{x}

\begin{document}

\title{Evidence for Black Hole Spin in GX 339-4:
  XMM-Newton EPIC-pn and RXTE Spectroscopy\\ of the Very High State}

\author{J.~M.~Miller\altaffilmark{1,2}, 
        A.~C.~Fabian\altaffilmark{3},
	C.~S.~Reynolds\altaffilmark{4},
	M.~A.~Nowak\altaffilmark{5},
	J.~Homan\altaffilmark{5},
	M.~J.~Freyberg\altaffilmark{6},
	M.~Ehle\altaffilmark{7},
	T.~Belloni\altaffilmark{8},		
	R.~Wijnands\altaffilmark{9},
	M.~van~der~Klis\altaffilmark{9},
	P.~A.~Charles\altaffilmark{10},
	W.~H.~G.~Lewin\altaffilmark{5},	
	}

\altaffiltext{1}{Harvard-Smithsonian Center for Astrophysics, 60
	Garden Street, Cambridge, MA 02138, jmmiller@cfa.harvard.edu}
\altaffiltext{2}{NSF Astronomy and Astrophysics Fellow}
\altaffiltext{3}{Institute of Astronomy, University of Cambridge,
        Madingley Road, Cambridge CB3 OHA, England, UK}
\altaffiltext{4}{Department of Astronomy, University of Maryland,
        College Park, MD, 20742}
\altaffiltext{5}{Center~for~Space~Research and Department~of~Physics,
        Massachusetts~Institute~of~Technology, Cambridge, MA
        02139--4307}
\altaffiltext{6}{Max-Planck-Institut f\"ur Extraterrestriche Physik,
        Giessenbachstr., D-85748 Garching, DE}
\altaffiltext{7}{\textit{XMM-Newton} SOC, Villafranca Satellite
        Tracking Station, PO Box 50727, 28080, Madrid, ES \& Research
        and Scientific Support Dept. of ESA, Noordwijk, NL}
\altaffiltext{8}{INAF -- Osservatorio Astronomico di Brera, Via
        E. Bianchi 46, I-3807, Merate, IT}
\altaffiltext{9}{Astronomical Institute ``Anton Pannekoek,''
        University of Amsterdam, and Center for High Energy
        Astrophysics, Kruislaan 403, 1098 SJ, Amsterdam, NL}
\altaffiltext{11}{Department of Physics and Astronomy, University of
        Southampton, SO17 1BJ, England, UK}

       
\authoremail{jmmiller@cfa.harvard.edu}

\label{firstpage}

\begin{abstract}
We have analyzed spectra of the Galactic black hole GX 339$-$4
obtained through simultaneous 76~ksec {\it XMM-Newton}/EPIC-pn and
10~ksec {\it RXTE} observations during a bright phase of its 2002--2003
outburst.  An extremely skewed, relativistic Fe~K$\alpha$ emission
line and ionized disk reflection spectrum are revealed in these
spectra.  Self-consistent models for the Fe~K$\alpha$ emission line
profile and disk reflection spectrum rule-out an inner disk radius
compatible with a Schwarzschild black hole at more than the 8$\sigma$
level of confidence.  The best-fit inner disk radius of $2-3~r_{g}$
suggests that GX~339$-$4 harbors a black hole with $a \geq 0.8-0.9$~
(where $r_{g} = GM/c^{2}$ and $a=cJ/GM^{2}$, and assuming that
reflection in the plunging region is relatively small).  This confirms
indications for black hole spin based on a {\it Chandra} spectrum
obtained later in the outburst.  The emission line and reflection
spectrum also rule-out a standard power-law disk emissivity in
GX~339$-$4; a broken power-law form with enhanced emissivity inside
$\sim6~r_{g}$ gives improved fits at more than the 8$\sigma$ level of
confidence.  The extreme red wing of the line and steep emissivity
require a centrally--concentrated source of hard X-rays which can
strongly illuminate the inner disk.  Hard X-ray emission from the base
of a jet --- enhanced by gravitational light bending effects --- could
create the concentrated hard X-ray emission; this process may be
related to magnetic connections between the black hole and the inner
disk.  We discuss these results within the context of recent results
from analyses of XTE~J1650$-$500 and MCG--6--30--15, and models for
the inner accretion flow environment around black holes.
\end{abstract}

\section{Introduction}
Irradiation of an accretion disk orbiting a black hole by a source of
hard X-rays can produce a fluorescent Fe~K$\alpha$ emission line,
which should bear the signatures of the strong Doppler shifts and
gravitational redshifts (Fabian et al.\ 1989; see also George \&
Fabian 1991).  If the black hole has near-maximal spin ($a \simeq
0.998$, where $a = cJ/GM^{2}$), the innermost stable circular orbit
(ISCO) around the black hole can be as small as $r_{in} = 1.24~r_{g}$
(where $r_{g} = GM/c^{2}$, note $r_{in} = 6~r_{g}$ for $a=0$); this
proximity is expected to produce Fe~K$\alpha$ emission line profiles
with strong red wings because of the relative importance of
gravitational red-shifts in comparison to Doppler shifts (Laor 1991).

In the X-ray spectra of supermassive black holes in AGN and
stellar-mass black holes in Galactic black hole candidates, skewed
Fe~K$\alpha$ line profiles have proved to be extremely important
diagnostics of the innermost relativistic regime (for AGN, see, e.g.,
Tanaka et al.\ 1995; for BHCs, see, e.g., Miller et al. 2002a; for a
review see Reynolds \& Nowak 2003).  In some cases, evidence for black
hole spin may be inferred by the line shape (for AGN, see, e.g.,
Iwasawa et al.\ 1999, Wilms et al.\ 2001, Fabian et al.\ 2003; for
BHCs see, e.g., Miller et al.\ 2002b, Miller et al.\
2004, Miniuti, Fabian, \& Miller 2004).

GX 339$-$4 is a recurrent, dynamically-constrained BHC ($M_{BH} \geq
5.8~M_{\odot}$; Hynes et al.\ 2003) in which radio jets with $v/c >
0.9$ have recently been observed (Gallo et al.\ 2004).  Herein, we
report on the time-averaged 76~ksec {\it XMM-Newton}/EPIC-pn and
10~ksec {\it RXTE} spectra of GX~339$-$4, obtained during a bright
phase (near 1~Crab in soft X-rays) of its 2002--2003 outburst.  

\section{Observation and Data Reduction}
GX~339$-$4 was observed with {\it XMM-Newton} for 75.6~ksec, starting
on 29 September 2002 09:06:42 UT (revolution 514).  The EPIC-pn camera
(Str\"uder et al.\ 2001) was operated in ``burst'' mode to accommodate
the high count rate expected during this observation.  The ``thin''
optical blocking filter was used.  The data were reduced using the
{\it XMM-Newton} suite SAS version 5.4.1, and the guidelines described
in the MPE ``cookbook'' (see
http://wave.xray.mpe.mpg.de/xmm/cookbook).  Events were extracted in a
stripe in RAWX (31.5--40.5) versus RAWY (2.5--178.5) space.   Due to the
extremely high source flux background events were not extracted.  The
events were then filtered by requiring ``FLAG=0'' (to reject bad
pixels and events too close to chip edges) and ``PATTERN $\leq$ 4''
(to accept singles and doubles), and the spectral channels were
grouped by a factor of 5 to create a spectrum.  The appropriate canned
burst mode response file was used to fit the spectrum.

{\it RXTE} observed GX~339$-$4 for 9.6~ksec starting on 29 September
2002 09:12:11:28.  The data were reduced using the suite LHEASOFT
version 5.2.  Standard time filtering (primarily, filtering-out the
SAA) returned net PCA and HEXTE exposures of 9.3~ksec and 3.3~ksec,
respectively.  For this analysis, we have only made use of the spectra
from PCU-2 (the best-calibrated PCU at the time of writing) and
HEXTE-A.  Events from all layers of PCU-2 were combined to make
spectra; ``pcabackest'' and the bright source background model were
used to make background spectra.  We added 0.75\% systematic errors to
the spectrum from PCU-2 using the tool ``grppha''.  A response matrix
(combining rmf and arf files) was generated using ``pcarsp''.  The
HEXTE spectral files were made using the standard recipes, and the
standard canned responses were used to fit the data.  The PCU-2
spectrum was fit in the 2.8--25.0~keV band (standard for {\it RXTE}
analysis; see, e.g., Park et al.\ 2003), and the HEXTE spectrum was
fit on the 20.0--100.0~keV band (the upper bound was fixed by the
upper bound over which the ionized disk reflection model is valid).
These {\it RXTE} spectra were fit jointly with an overall normalizing
constant allowed to float between them.  An edge was added at 4.78~keV
with $\tau=0.1$ to correct for an instrumental Xe edge at this energy
in fits to the PCA spectrum.

\section{Analysis and Results}

Fits to the {\it XMM-Newton}/EPIC-pn spectrum, and joint fits to the
EPIC-pn and {\it RXTE} spectra do not yield formally acceptable fits
for the models discussed below.  A number of sharp, narrow calibration
uncertainties remain in the EPIC-pn detector response.  For the most
part, these appear as absorption lines or edges in the 2--3~keV range
(a feature at 2.31~keV appears as an emission line), though similar
features are present in the 0.7--2.0~keV band.  Most of these features
are likely due to Au M-shell edges and Si features in the detector;
they are revealed clearly in this observation due to the high
signal-to-noise achieved.  These narrow-band features do not affect
measurements of the Fe~K$\alpha$ emission line profile, but they do
affect the overall $\chi^{2}$ fit statistic.  Due to these
complications, within this analysis we use the broad-band capacity of
{\it RXTE} to characterize the continuum to serve as a guide for
fitting the EPIC-pn data, and reserve detailed joint fits for future
work.  All fits were made using XSPEC version 11.2.0 (Arnaud 1996).
All errors in this work are 90\% confidence errors.

The ``Laor'' model (Laor 1991) describes the line profiles expected
around a Kerr black hole.  The parameters of this model are the line
energy $E_{Laor}$, the disk emissivity index $q$ (where the emissivity
is assumed to have a power-law form of $J(r) \propto r^{-q}$; $q=3$ is
expected for a standard disk), the inner radius of the line emission
region $r_{in}$ in units of $r_{g} = GM/c^{2}$ (with a lower limit of
$r_{in} = 1.235~r_{g}$ for $a=0.998$), the outer line emission radius
(fixed to the model limit of $R_{out} = 400~r_{g}$), the disk
inclination, and the line normalization.

The ``pexriv'' reflection model (Magdziarz \& Zdziarksi 1995)
describes reflection from an ionized accretion disk, and does not
explicitly include Fe~K$\alpha$ line emission.  The important
parameters in this model are the reflection fraction $f$ ($f =
\Omega/2\pi$), 

\centerline{~\psfig{file=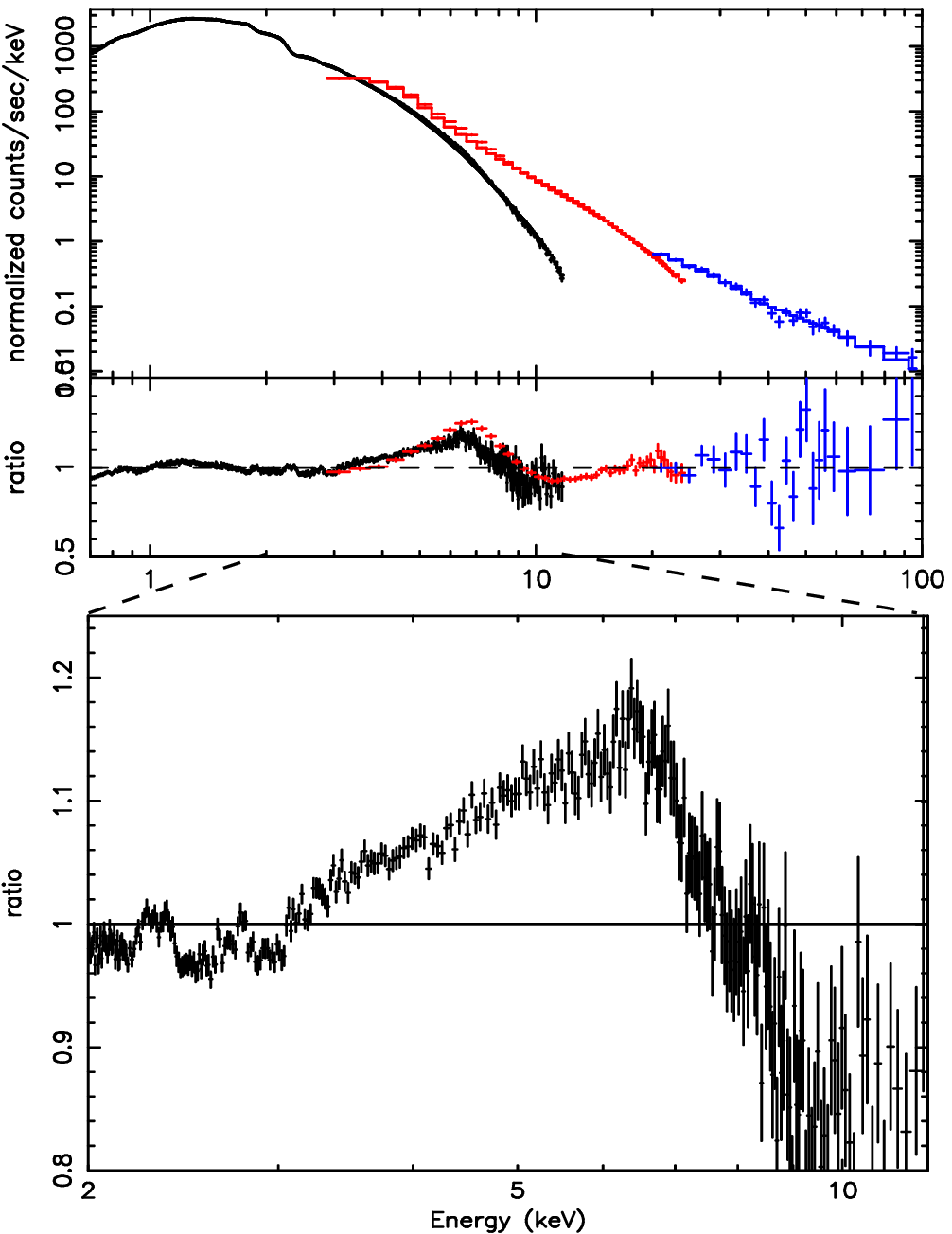,width=3.2in}~}
\figcaption[h]{\footnotesize Above: the data/model ratio obtained when
the {\it XMM-Newton}/EPIC-pn spectrum (black) and {\it RXTE}/PCA (red)
and HEXTE (blue) spectra are jointly fit with a model consisting of
MCD and power-law components.  The 4.0--7.0~keV range was ignored when
fitting the data, and the spectra were re-binned for clarity.
Below: a blow-up of the EPIC-pn spectrum showing the extremely skewed
line profile.}
\medskip

the disk temperature, ionization ($\xi = L_{X}/nr^{2}$), and
inclination, the power-law folding energy ($E_{fold}$), and the index
and normalization of the power-law flux irradiating the disk.  The
``constant density ionized disk'' reflection model (CDID; Ross,
Fabian, \& Young 1999; Ballantyne, Iwasawa, \& Fabian 2001) is a
reflection model which includes Fe~K$\alpha$ line emission, and is
especially well-suited to high ionization regimes.  The relevant
parameters of this model are ${\rm log}(\xi)$, $f$, and the index and
normalization of the power-law flux irradiating the disk.

These disk reflection spectra are calculated in the co-rotating or
fluid frame, and so must be convolved (or, ``blurred'') with the Laor
line element describing the Doppler shifts and gravitational
red-shifts expected for an observer in a stationary frame at infinity.
In blurring the reflected spectra (and, therefore, the power-law
continuum), we linked the parameters of the emission line and blurring
function.  The disk components were not blurred as the multicolor disk
blackbody model (MCD; Mitsuda et al.\ 1984) --- only an approximation
to the Shakura \& Sunyaev (1973) disk model because it lacks an inner
boundary term --- may be the best--available model for a disk orbiting
a spinning black hole when torques are present at the inner boundary.

\subsection{Fits to the RXTE Spectra}

In fits to the {\it RXTE} spectra, the equivalent neutral hydrogen
column density was fixed at $N_{H} = 5.3\times 10^{21}~{\rm cm}^{-2}$
(Dickey \& Lockman 1990) using the ``phabs'' model.  No reasonable
continuum model gives an acceptable fit to the {\it RXTE} spectra
without the addition of a broad Fe~K$\alpha$ emission line.  In the
absence of added line components, the best-fit ``canonical'' MCD plus
power-law model gives $\chi^{2}/\nu = 279.8/74$, the bulk-motion
Comptonization model (Shrader \& Titarchuk 1999) gives $\chi^{2}/\nu =
766.1/74$, and an MCD plus CompTT (Titarchuk 1994) continuum gives
$\chi^{2}/\nu = 284.9/72$ (where $\nu$ is the number of degrees of
freedom).  Adding Gaussian emission line and smeared edge (``smedge'')
components to the MCD plus power-law model as an approximation to a
reflection model, a good fit is achieved ( $\chi^{2}/\nu = 68.2/68$).
The parameters measured via this model are: $kT =
0.84^{+0.06}_{-0.04}$~keV, $K_{MCD} = 2800 \pm 600$, $\Gamma =
2.5^{+0.2}_{-0.1}$, $K_{pl} = 2.6 \pm 0.4$, $E_{Gauss} =
5.8^{+0.6}_{-0.8}$~keV, FWHM$ = 3^{+1}_{-2}$~keV, $K_{Gauss} =
3_{-2}^{+3} \times 10^{-2}$, EW$ = 300^{+300}_{-100}$~eV, $E_{smedge}
= 8.7_{-0.9}^{+0.6}$~keV, $\tau = 0.2_{-0.2}^{+0.8}$, and $W_{sm} =
2_{-1}^{+2}$~keV (where $K$ denotes model normalizations, and the
``smedge'' energy was constrained to lie in the 7.1--9.3~keV range).

Fits were also made with blurred CDID and pexriv models.  Blurring the
models for the $1.24-400~r_{g}$ range and assuming a power-law
emissivity of $q=3.0$, we found that the data are consistent with
strong reflection ($f \simeq 1$) from a highly ionized inner disk
(${\rm log}(\xi) = 4.3-4.4$).  A line equivalent width of $EW \simeq
200\pm 80$~eV was meausured using the pexriv model.

\subsection{Fits to the XMM-Newton/EPIC-pn Spectrum}

Due to the hints of an extremely skewed spectrum found with {\it
RXTE}, we began by fitting more physical models to the {\it
XMM-Newton}/EPIC-pn spectrum.  Broad line residuals extend down to
$\sim$3~keV regardless of whether models with additive components or
single-component (e.g., bulk motion Comptonization) models are used to
fit the continuum, which indicates that the continuum does not
strongly affect the red wing of the line.  The first model we
considered included MCD and power-law continuum components, with added
Laor line and smeared edge components.  The power-law index was
constrained to lie within $\Delta(\Gamma) \leq 0.1$ of the {\it
RXTE}-measured values.  With this model, we measure the following
values: $N_{H} = 5.1\pm 0.1 \times 10^{21}~{\rm cm}^{-2}$, $kT =
0.76\pm 0.01$~keV, $K_{MCD} = 2300^{+100}_{-200}$, $\Gamma =
2.60_{-0.05}$, $K_{pl} = 2.2_{-0.1}^{+0.3}$, $E_{Laor} =
6.97_{-0.20}$~keV, $q = 5.5_{-0.1}^{+0.5}$, $r_{in} =
2.1^{+0.2}_{-0.1}~r_{g}$, $i = 11^{+5}_{-1}$~deg., $K_{laor} =
7.7_{-1.5}^{+0.5} \times 10^{-2}$, $EW_{Laor} = 200^{+10}_{-40}$~eV,
$E_{sm} = 7.9_{-0.4}^{+0.1}$~keV, $\tau = 0.6_{-0.1}^{+0.4}$, and a
smeared edge width of $W = 1.0 \pm 0.3$~keV.  An inner radius of
$r_{in} = 6~r_{g}$ (as per a black hole with $a=0$) is excluded at
more than the 8$\sigma$ level of confidence, as is a standard
emissivity index of $q=3$.  This model gives $\chi^{2}/\nu =
3456.5/1894$; the formally unacceptable fit is due to the calibration
problems discussed in Section 3.0.  A joint fit to the {\it
XMM-Newton}/EPIC-pn and {\it RXTE} spectra using these continuum
parameters is shown in Figure 1.  These spectral parameters, and the
timing properties of the source at the time of this observation (Homan
et al.\ 2004), indicate that GX~339$-$4 was observed in the ``very
high'' state.

Using this model, we measure an unabsorbed flux of $2.1\times 10^{-8}~
{\rm erg}~ {\rm cm}^{-2}~ {\rm s}^{-1}$ in the 0.5--10.0~keV band; the
power-law contributes 35\% of this flux.  The total line flux is
measured to be $5.3\times 10^{-10}~ {\rm erg}~ {\rm cm}^{-2}~
{\rm s}^{-1}$, or $8.4\times 10^{-2}~{\rm ph}~{\rm cm}^{-2}~{\rm s}$.
It should be noted that the best fit without a line component gives
$\chi^{2}/\nu = 7552/1902$.  Clearly, a broad line component is
required at far more than the $8\sigma$ level of confidence.

Models which allow the disk emissivity law to assume a broken
power-law form, with $J(r) \propto r^{-q_{in}}$ within radius $r_{q}$,
and $J(r) \propto r^{-q_{out}}$ outside of $r_{q}$, yield
statistically improved fits over models assuming simple power-law
emissivity laws at more than the $8\sigma$ level of confidence.

Allowing for a broken power-law form for the emissivity and a blurred
CDID reflection model, we measure: $N_{H} = 5.5\pm 0.1 \times
10^{21}~{\rm cm}^{-2}$, $kT = 0.79\pm 0.01$~keV, $K_{MCD} = 1860\pm
20$, 

\centerline{~\psfig{file=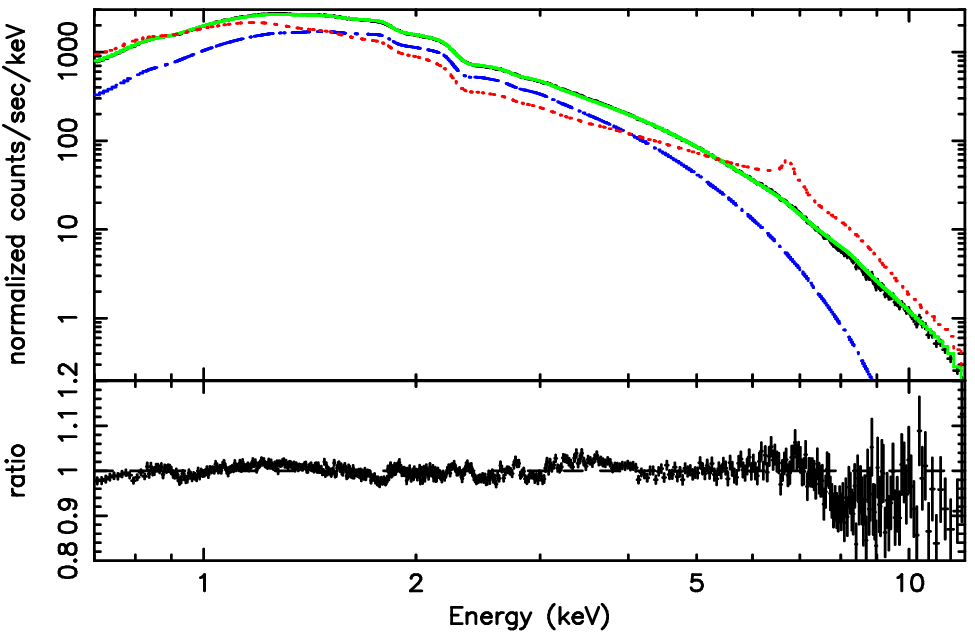,width=3.2in}~}
\figcaption[h]{\footnotesize The EPIC-pn spectrum of GX~339$-$4 fit
with a relativistically blurred reflection model, with data/model
ratio shown below.  The total spectrum is shown in green, the disk
component in blue, and the CDID reflection model is shown in red
(prior to blurring).  The data were re-binned for clarity.}
\medskip

$\Gamma = 2.61_{-0.01}^{+0.09}$, $K_{CDID} = 2.8\pm 0.2 \times
10^{-26}$, ${\rm log}(\xi) = 4.5_{-0.1}^{+0.5}$, $f = 2.0_{-0.2}$,
$r_{in} = 2.9 \pm 0.1$, $q_{in} = 5.5^{+0.5}_{-0.7}$, $q_{out} =
3.0^{+0.6}$, $r_{q} = 7\pm 2~r_{g}$, and $i = 11_{-1}^{+5}$~deg., for
$\chi^{2}/\nu = 3887.3/1895$.  This fit to the EPIC-pn spectrum is
shown in Figure 2.

Fits with a blurred pexriv model with a broken power-law emissivity
($\xi = 3.0\times 10^{4}$, $kT_{disk} = 1.0$~keV as per Young et al.\
2001, $E_{fold} = 200$~keV, $r_{q} = 6~r_{g}$, and $q_{out} = 3$ were
fixed) yielded the following fit parameters: $N_{H} = 5.3\pm 0.1
\times 10^{21}~{\rm cm}^{-2}$, $kT = 0.76\pm 0.01$~keV, $K_{MCD} =
2200\pm 300$, $\Gamma = 2.61_{-0.01}^{+0.09}$, $K_{PL,pexriv} = 1.5\pm
0.1$, $i = 12_{-2}^{+4}$~deg., $E_{laor} = 6.97_{-0.07}$~keV, $r_{in}
= 2.1_{-0.3}^{+0.5}$, $q_{in} = 5.5_{-0.2}^{+0.5}$, $K_{laor} = 7.2\pm
0.3 \times 10^{-2}$, $EW = 210^{+60}_{-60}$~eV, and $f =
1.0^{+0.5}_{-0.1}$, giving $\chi^{2}/\nu = 3481.3/1896$. 

The measured inner disk color temperature and high disk ionizations
are expected in the very high state.  Reflection fractions greater
than unity might be taken as indications of light bending effects.
All fits show negative residuals above $E \simeq 8.0-8.5$~keV.
Reflection models predict a steeper continuum and also account for
ionized Fe~K-shell absorption in this range.  The observed residuals
can be corrected with a phenomenological smeared edge component
``$\tau$''~$\simeq 0.5$,but would be better modeled by refinements to
reflection models.

At high mass accretion rates, disks around black holes likely extend
to the ISCO.  Although the Laor model assumes maximal black hole spin,
it is likely broadly valid over a range of high spin parameters.
Assuming that $r_{in} = r_{ISCO}$ (equivalent to assuming that the
inner disk drags matter in the plunging region minimally; see Krolik
\& Hawley 2002, Reynolds \& Begelman 1997, and Young, Ross, \&
Fabian 1998); equation 2.21 in Bardeen, Press, \& Teukolsky (1972) may
be used to estimate the spin parameter of the black hole in
GX~339$-$4.  Using this procedure, the line and reflection fits
detailed above suggest that GX~339$-$4 harbors a black hole with $a
\geq 0.8-0.9$.

\section{Discussion}
We have discovered an extremely skewed Fe~K$\alpha$ emission line in
an {\it XMM-Newton}/EPIC-pn spectrum of the Galactic black hole
GX~339$-$4.  Spectral fits with the Laor relativistic disk line model
and a phenomenological continuum, and with the Laor model and
self-consistent blurred disk reflection models, provide tight
constraints on the nature of the black hole and inner accretion flow
in GX~339$-$4.  Our fits indicate that the inner disk likely extends
to $r_{in} = 2-3~r_{g}$, translating to a black hole spin parameter of
$a \geq 0.8-0.9$.  Moreover, an enhanced inner disk emissivity index
of $q=4.8-6.0$ is required within $5-9~r_{g}$.  Fits with the same
models which fix either $r_{in} = 6.0~r_{g}$ (as per an $a=0$
Schwarzschild hole) or $q=3.0$ (as per a standard disk), are more than
8$\sigma$ worse than the best-fit values summarized here and detailed
in the previous section.  

X-ray spectroscopy of XTE J1650$-$500 (Miller et al.\ 2002b, Miniutti,
Fabian, \& Miller 2003) and MCG--6-30-15 (Wilms et al.\ 2001, Fabian
et al.\ 2002) has revealed remarkably similar line profiles and inner
disk emissivity properties; both XTE~J1650$-$500 and MCG--6-30-15 may
have spin parameters which are similar to GX~339$-$4.  Comparisons of
the timing phenomena observed in stellar-mass Galactic black holes and
Seyfert galaxies have shown that noise properties may simply scale
with mass (Uttley \& McHardy 2001).  It has also recently been shown
that Galactic black holes may have AGN-like warm absorbers (Miller et
al.\ 2004).  Our results extend these findings in that they suggest
that --- at least in certain states --- details like enhanced inner
disk emissivity are likely similar.

The models we have considered not only constrain the black hole spin
parameter, but also the nature of the inner accretion flow geometry.
The inner disk must be strongly illuminated by a
centrally-concentrated source of hard X-rays.  Miniutti \& Fabian
(2003) have calculated the effects of light bending on Fe~K$\alpha$
line variability.  This model can explain the complex time variability
seen in MCG--6-30-15, and appears to describe the behavior of the line
strength in XTE~J1650$-$500 remarkably well (Miniutti, Fabian, \&
Miller 2004; Rossi et al.\ 2003).  The light bending model assumes a
power-law source of hard X-ray emission above the black hole which
moves ``vertically'' along the black hole and inner disk angular
momentum axis.  While models for jets in Galactic black holes require
that X-ray emission be focused away from the disk (see, e.g., Markoff,
Falcke, \& Fender 2001), at the base of the jet the flow is less
relativistic and the beaming therefore less extreme.  It is possible
that synchrotron self-Compton emission from the base of a jet may be
focused onto the inner disk by light bending (this would represent an
extension to the jet reflection considered by Markoff \& Nowak 2004).
This does not rule-out additional hard X-ray emission from a corona.

Heightened inner disk emissivity of the kind seen in GX~339$-$4 can
plausibly be explained by the dissipation of the black hole's
rotational energy via magnetic connections to the inner disk
(Blandford \& Znajek 1977; see also Gammie 1999 and Li 2003).  The
emissivity we have measured is above that predicted by magnetic
connections to matter in the plunging region ($q = 7/2$, Agol \&
Krolik 2000), though it is possible this process could also be at
work.  It is possible that such processes and the putative
illumination of the inner disk region by the base of a jet might be
intimately related; indeed, the extraction of black hole spin energy
is often invoked as a possible means of powering relativistic jets in
AGN.  

We are grateful for thoughtful comments of the anonymous referee.  JMM
gratefully acknowledges support from the NSF through its Astronomy and
Astrophysics Postdoctoral Fellowship program.  This work is based on
observations obtained with \textit{XMM-Newton}, an ESA science mission
with instruments and contributions directly funded by ESA Member
States and the USA (NASA).

\end{document}